# Title

µSpikeHunter: An advanced computational tool for the analysis of neuronal communication and action potential propagation in microfluidic platforms

## Authors

Kristine Heiney, José Mateus, Cátia Lopes, Estrela Neto, Meriem Lamghari, Paulo Aguiar*

## Affiliations

i3S – Instituto de Investigação e Inovação em Saúde, Universidade do Porto, Rua Alfredo Allen, 208, 4200-135 Porto, Portugal

INEB – Instituto de Engenharia Biomédica, Universidade do Porto, Rua Alfredo Allen, 208, 4200-135 Porto, Portugal

Corresponding Author:
Paulo de Castro Aguiar: pauloaguiar@ineb.up.pt

# Abstract

Understanding neuronal communication is fundamental in neuroscience but there are few methodologies offering detailed analysis for well-controlled conditions. By interfacing microElectrode arrays with microFluidics (μEF devices), it is possible to compartmentalize neuronal cultures with a specified alignment of axons and microelectrodes. This setup allows extracellular recordings of spike propagation with high signal-to-noise ratio over the course of several weeks. Addressing these μEF systems we developed an advanced, yet easy-to-use, open-source computational tool, μSpikeHunter, which provides detailed quantification of several communication-related properties such as propagation velocity, conduction failure, spike timings, and coding mechanisms. The combination of μEF devices and μSpikeHunter can be used in the context of standard neuronal cultures or with co-culture configurations where, for example, communication between sensory neurons and other cell types is monitored and assessed. The ability to analyze axonal signals (in a user-friendly, time-efficient, high-throughput manner) opens doors to new approaches to studies of peripheral innervation, neural coding, and neuroregeneration approaches, among many others. We demonstrate the use of μSpikeHunter in dorsal root ganglion neurons where we analyze the presence of anterograde signals in μEF devices.

# Introduction

Electrical signaling is recognized as the principal modality of communication in neurons, where it is used to encode and transmit information via action potentials (APs). Electrophysiology recordings play therefore a fundamental role in understanding neuronal circuits in physiological and pathological conditions. Multielectrode arrays (MEAs) standout among the different methodologies with the appropriate spatial and temporal scales to assess neuronal circuits in well-controlled in vitro settings [1,2]. However, it is difficult to detect and characterize AP propagation (e.g., direction and velocity) using conventional MEAs in neuronal cultures: it is virtually impossible to ensure that electrodes are well placed for APs detection, source-target information is inaccessible, and the amplitude of the APs recorded from axons are typically very low and impossible to discriminate. To improve the signal-to-noise ratio (SNR) of electrophysiological recordings and the localization of neuronal processes on recording electrodes, devices combining microElectrodes and microFluidics (µEF devices) have been developed for neuroscience applications [3]. A µEF device is composed of a microfluidic device mounted on a MEA to form an enclosed culture chamber composed of two (or more) isolated compartments connected by microchannels. The reduced dimensions of the microchannels ensure that somata are excluded from these channels whereas axons are able to grow through. Among other applications, this type of device allows for the spatially compartmentalized but functionally connected cultures of separate populations of cells. With the microfluidic device aligned such that the microchannels are positioned above multiple microelectrodes, signal propagation along the axons can be observed and analyzed. The small dimensions of the microchannels provide the added benefit of increasing the SNR of recorded axonal APs [4].

A number of studies have been conducted using µEF devices to assess i) the directionality of communication and the origin of bursting behavior in networks of distinct populations of neurons [5–9], ii) changes to the propagation velocity with culture age [10,11], or iii) results of pharmacological, biochemical, or electrical stimulation [10,12,13].

Compartmentalized microfluidic devices have also been used to investigate the interaction between neurons and other cells co-cultured in the separate compartments. But although studies on neuronal circuit dynamics and signal communication are intimately related with electrophysiology, the vast majority of these microfluidic studies on neuronal co-cultures with other types of cells have lacked an electrophysiological facet, which would serve to complement the fundamentally biochemical results and further elucidate the interaction between the cell types. For example, past studies have assessed various factors affecting myelination in the central nervous system [14–16]. Additionally, the biochemical and morphological facets of the interaction between neurons and cells from other organ systems, including osteoblasts [17], dental pulp [18], and myocytes [19,20] have been investigated.

In both single- and co-cultures configurations, more than the technical difficulty of combining microelectrodes and microfluidics, the challenge for most labs lies on the sheer volume and complexity of the recorded electrophysiology data. The electrophysiology data obtained from neurons cultured in µEF devices is unquestionably highly informative, allowing for example for the detection of propagating APs, calculation of their propagation direction and velocity as well as their spike times or inter-spikes interval (relevant for neuronal coding/decoding analysis). However, although the necessary experimental protocols and tools for combining MEAs and microfluidics are readily available [JoVE REF], there is a lack of user-friendly tools available to analyze recordings obtained using µEF devices (but see [21,22], for visualization and spike sorting tools).

In this study, an advanced yet user-friendly data analysis program called µSpikeHunter was developed for the detection and characterization of APs propagating along axons confined to the microchannels of a µEF device. With this free, open-source software, the presence of traveling waves (APs) in electrophysiological recordings can be readily detected, visualized and characterized. Spike sorting can also be performed in µSpikeHunter based on the waveforms of recorded APs. In this study, µSpikeHunter was validated in data analysis of electrophysiology experiments using cortical neurons and dorsal root ganglia (DRGs), and was used to uncover the otherwise elusive presence of APs traveling towards the somata compartment in common microfluidic devices.

# Results

## µSpikeHunter graphical user interfaces

The computational tool µSpikeHunter runs on both *Microsoft Windows* and *Apple's macOS* operating systems, and is composed of two graphical user interfaces (GUIs): the main GUI, in which the data are imported and analyzed at the single-spike level, and the spike sorting GUI, in which the user can sort spikes into source clusters associated to different neurites. µSpikeHunter was developed to be compatible with recordings obtained using custom setups or commercial recording systems (MEA2100 from MultiChannel Systems) for 60-, 120-, and 252-electrode MEAs. Analysis can be performed for signals recorded by a series of up to 16 electrodes with a uniform inter-electrode spacing. Details on how to use µSpikeHunter can be found in the detailed user manual in the supplementary materials.

The main GUI (Fig. 1) allows the user to select a file for analysis and, if a commercial MultiChannel Systems setup was used for the recording, the electrodes and time range for analysis. Once the desired data have been selected and the parameters set for event detection, the user is presented with a list of detected propagation sequences. These sequences are series of events detected along the electrodes which satisfy specific criteria (see Materials and Methods) to be considered an AP traveling (in an axon) through the microchannel containing the analyzed electrodes. The propagation sequences can be analyzed/characterized and the results exported to a data file.

The spike sorting GUI (Fig. 2) presents the user with a spike overlay for each of the electrodes consisting of all the detected propagation sequences, and allows the user to sort the spikes into source clusters, which are collections of propagation sequences that are presumed to have arisen from different axons inside the same microchannel. The spike overlay plots presented to the user are aligned about the peak voltage values for each detected event.

## Algorithms performance assessment using synthetic data

The µSpikeHunter algorithms for detection and characterization of propagating APs were assessed/validated using synthetic data simulating electrophysiological recordings of microelectrodes along a microchannel, with different levels of SNR.

*Propagation sequence detection*

The propagating wave detector was evaluated based on two performance indices: the precision rate and the detection rate. The precision is defined as

$$PR = \frac{TP}{TP + FP}$$

where $TP$ is the number of true positives, i.e., the number of detected sequences that correspond to actual traveling waves, and $FP$ is the number of false positives, i.e., the number of detected sequences that do not correspond to actual traveling waves. The detection rate is defined as

$$DR = \frac{TP}{NS}$$

where $NS$ is the total number of actual sequences in the recording dataset. Both of these performance indices range from 0 to 1.

A high precision indicates that the propagation sequence detector rarely yields temporally linked events that do not correspond to actual traveling waves (APs). A high detection rate indicates that the propagation sequence detector is able to recognize a high percentage of the actual traveling waves in the data. It should be noted that in spike characterization, a high precision is more important than a high detection rate. That is, it is more desirable to be certain that the analyzed spikes are actually traveling APs than to detect the majority of spikes in a recording (due to noise and detection constrains, the same spike may not be observed in all microelectrodes along a microchannel).

The precision and detection rate results are presented in Fig. 3 for different levels of SNRs. Examples of generated propagation sequences with no added noise and with added noise at SNRs of 0.7 and 0.3 are show in Fig. 3a. The plotted values for precision and detection rates (Fig. 3b) are the averages of the values obtained for multiple simulations using the same SNR. These results demonstrate that the propagation sequence detector shows very high precision up to an SNR of 0.3 (significantly below typical values for standard good quality MEA recordings), where it is still very close to 1 (0.96).

The detection rate drops steadily as the noise increases from a SNR of 0.7 to 0.2. Whereas the propagation sequence detector is able to detect 83% of all actual propagation sequences at a SNR of 0.7, for a SNR of 0.4 less than 10% of all propagation sequences are detected.

*Propagation velocity estimation*

The performance of the propagation velocity estimation of µSpikeHunter was assessed using three measures (see Materials and Methods):

- cluster propagation velocity (CPV), calculated from the relative timing and shape of the voltage waveforms, measured between the two most distant electrodes;
- single-sequence propagation velocity (SPV), calculated directly from the cross-correlation between the voltage waveforms in a particular pair of electrodes;
- mean SPV, calculated using the average SPV for all electrode pairs.

A comparison of the methods is shown in Fig. 3c. These results indicate the average propagation velocity estimates over all detected propagation sequences in the synthetic datasets. All methods show a good performance at high SNR, with SPV between specific pairs of electrodes offering a versatile tool. At low SNR however, CPV outperforms the other methods and shows good performance up to an SNR of 0.4, with an average error of 2.2%. Beyond a SNR of 0.3, the performance of the CPV deteriorates.

**Detailed characterization of axonal signal propagation in controlled *in vitro* settings**

Two sets of experiments, using microfluidic platforms with a somal and an axonal compartment separated by microchannels, were used to demonstrate the data analysis capabilities of µSpikeHunter.

*Spike sorting*

The spike sorting performance of µSpikeHunter was evaluated using recordings obtained from rat cortical neurons at DIV 15 (days *in vitro*) using MEAs with 252 microelectrodes and a sampling rate of 20 kHz. Propagation sequences were detected with a threshold of 5 standard deviations of the noise and negative phase detection. The presence of distinct spike waveforms in the spike sorting GUI is a strong indication that multiple sources of activity are present in the microchannels. The sorting results for two microchannels (B and M; see Fig. 1 for a schematic example of this labeling system) of one of the experiments are shown in Fig. 4a1 and b1. The characteristics of the sorted clusters are summarized in Fig. 4a2 and b2. µSpikeHunter's spike sorting GUI provides the number (and timings) of the spikes in each cluster and their propagation velocities.

Interestingly, in addition to the distinct appearance of the waveforms, the propagation velocities of the sorted clusters can provide further confirmation that the spikes arise from different sources. For example, in Fig. 4a2 and b2 clusters 1 and 2 in channel B show significantly different CPVs with low standard deviations corresponding to errors of approximately 5% and high confidence indices owing to the highly consistent waveforms in each cluster. The same high reliability can also be observed in the CPV estimate for cluster 1 in channel M, which has an error of approximately 8% at one standard deviation. However, the standard deviation of the CPV estimates were higher for cluster 0 in both channels B and M; in these cases, the error exceeded 20%. This is likely because the SNRs of the spikes in these clusters were lower and the propagation velocity estimates were therefore more affected by noise. Alternatively, it may also be the case that these clusters actually represent more than one source, but the waveforms are indistinguishable, again because of the SNR.

*Observation of reverse propagation in standard microfluidics with somal/axonal compartments*

In microfluidic experiments where neurons are placed only in the somal compartment it is expected that all APs travel from the somal to the axonal compartment (forward propagation). However, taking advantage of μSpikeHunter analysis and visualization capabilities, this is not what is observed. In experiments with either cortical neurons or DRG, a significant number of events (~10% in some cases) travel from the axonal to the somal compartment (reverse propagation). Examples of reverse propagation sequences are shown in Fig. 5, presented as voltage traces and kymographs. The kymographs in μSpikeHunter allow the user to visually, and readily, identify the occurrence of propagation (either forward or reverse).

Possible causes for the observed reverse propagation include (Fig. 6): distal axo-axonic synapses in the axonal compartment (between axons of different or same microchannel); axons growing back towards the somal compartment (through different or same microchannel); and antidromic propagation with the action potential being generated at the axon terminals.

To analyze the possible causes, the timings and velocities of the detected reverse propagation sequences in each recorded microchannels (and in each source cluster) were check against the forward sequences in all other recorded microchannels. In particular, the times at which a reverse APs was detected on the most distal electrodes (with respect to the somal compartment) was compared with APs detection time in other microchannels/source clusters. Strong correlation, short time delays, and low variability in the time delays are a good indication that an axon grew back from another microchannel or, alternatively, formed an excitatory axo-axonal synapse with a source in another microchannel (potential causal relation between APs in different microchannels/sources)

In cortical cultures it was possible to observe reverse APs correlated with forward APs in different microchannels/sources. In some cases, more than 90% of all detected reverse APs in a cluster were preceded by forward APs in a different cluster/microchannel, with delays between the APs arrival times (on distal electrodes) on the order of 1.8 ms ± 0.07 ms (mean ± std). In other cases it was not possible to relate detected reverse APs with any recorded APs in different sources/microchannels; however, the µEF device used in this study has 20 microchannels, and so it cannot be excluded that an axon from one of the four silent microchannels may have been the source of the reverse propagation. Nevertheless, sources (after sorting) showing reverse APs were never exclusive to reverse propagation: a source could should only forward APs or have up to a fraction of 0.9 reverse APs, but interestingly was never exclusive to reverse propagation. While not excluding the possibility of indistinct sources on the same microchannel, given the observed same velocities for forward and reverse propagation in these cases, this result raises the possibility of both orthodromic and antidromic conduction in the same axon in this experimental setup/conditions.

The DRG culture showed reverse propagation without the same regularity as the cortical culture. Few reverse propagating events were observed, and they were often isolated events in different microchannels and on different DIVs. This indicates, with increased confidence, that these events likely represent the antidromic propagation (in line with physiological conditions where DRG integrates afferent signals).

*Monitoring culture activity over multiple days in vitro*

As a final benchmark for the analysis capabilities µSpikeHunter a DRG explant culture was monitored in a µEF device as it aged, using 10 minutes recordings at DIVs 4, 6, and 8. Propagation sequences were detected using a threshold of 3.5 standard deviations of the noise with negative phase detection. The activity was monitored by determining the number of APs in each 10 min recording, the SNR, and the CPV estimate. The results are shown in Fig. 7, focusing on the evolution of two microchannels (K and H) that consistently showed high levels of activity from a single source. Figure 7a shows the propagation velocity and number of events at the three measurement time-points for the two considered activity sources. From DIV 4 to DIV 8, the number of events in channel K increased steadily in contrast with channel H. Furthermore, in channel K, the propagation velocity (related to axon caliber and ion channels density) was observed to increase from 0.57 to 0.77 m/s between DIVs 4 and 6 and then remain constant between DIVs 6 and 8, whereas in channel H, the propagation velocity initially remained roughly constant at approximately 0.65 m/s and then decreased to 0.46 m/s at DIV 8. Figure 7b shows the SNRs on each of the electrodes in channels K and H at the three measurement points. The SNR was highest on DIV 6 across all electrodes in both channels K and H. It was also possible to observe that a higher SNR is achieved closer to the center of the microchannels, as has been described previously [4].

Altogether, these results give an indication of how informative and useful µSpikeHunter is, and how it can allow researchers to easily monitor (co-)culture development, activity and communication over multiple DIVs.

# Discussion

µSpikeHunter is a powerful yet user-friendly program for the identification and characterization of propagating APs in neuronal (co-)cultures with a controlled compartmentalization. The developed algorithms in this computational tool allows the user to determine if there are APs propagating along axons confined to microchannels of µEF devices and characterize these APs based on the direction of propagation, the propagation velocity, firing rate, spike timings, the SNR, and the source. The AP detection and the performance of the propagation velocity estimation algorithms of µSpikeHunter were assessed/validated using synthetic APs generated with different levels of noise. It was found that the precision of the spike detection and the CPV estimate show good performance up to an SNR of 0.3, way below the typical values in good electrophysiological recordings. The program is robust, versatile and can be used with different neuronal cultures, such as cortical neurons or DRG explants, including co-cultures configurations. Altogether, the methods presented here combining µEF devices and µSpikeHunter provide very interesting conditions to study (in an easy, time-efficient, high-throughput manner), neurodevelopment, neuronal circuits, information coding, neurodegeneration and neuroregeneration, among others. The use of µEF devices also allows the same neurites to be monitored over multiple DIVs, and µSpikeHunter simplifies the analysis of the evolution of spike features in this type of analysis.

µSpikeHunter was used to uncover the otherwise elusive presence of APs traveling towards the somata compartment in common microfluidic devices. The observation of reverse propagation on the microchannels also provides insight into how µSpikeHunter can be used to assess the afferent response of neurons to changes in the culture conditions in the axonal compartment, or the communication between neurons and other cells in co-cultures.

µSpikeHunter is available for download from [SourceForge/Github links]

## Materials and Methods

### Preparation of microelectrode–microfluidic devices

The µEF devices were prepared by placing standard microfluidic devices with an appropriate microgrooves spacing against a pre-coated MEA (MultiChannel Systems MCS GmbH, Germany) with 120 or 252 electrodes of 30 µm in diameter with a center-to-center inter-electrode spacing of 100 µm. A photograph of a prepared µEF device is shown in (Supplementary Material) Fig. 8. Briefly, the MEAs were coated with 0.01 mg/ml of poly(D-lysine) (PDL, Corning) overnight at 37 °C, washed with sterile water, and completely air-dried under sterile conditions. Microfluidic devices were sterilized with 70% ethanol and were gently attached to PDL-coated MEAs, creating a µEF chamber composed of two separate compartments connected by 450 µm length × 9.6 µm height × 14 µm width microchannels. The medium reservoirs were loaded with 150 µl of 5 µg/ml laminin isolated from mouse Engelbreth-Holm-Swarm sarcoma (Sigma-Aldrich Co.) and incubated overnight at 37 °C. The unbound laminin-1 was removed, and chambers were refilled with Neurobasal medium and left to equilibrate for at least 2 h at 37 °C prior to cell seeding.

### Cell culture experiments

Experimental procedures involving animals were carried out in accordance with current Portuguese laws on Animal Care (DL 113/2013) and with the European Union Directive (2010/63/EU) on the protection of animals used for experimental and other scientific purposes. The experimental protocol (reference 0421/000/000/2017) was approved by the ethics committee of the Portuguese official authority on animal welfare and experimentation (Direção-Geral de Alimentação e Veterinária). All possible efforts were made to minimize the number of animals and their suffering. Unless otherwise stated, all reagents listed below are from Gibco, ThermoFisher Scientific.

Primary embryonic rat cortical neurons were isolated from prefrontal cortices of Wistar rat embryos (E18). The embryo cortices were dissected in Hank's Balanced Salt Solution (HBSS) and enzymatically digested in 0.6% trypsin (1:250) in HBSS for 15 min at 37 °C. Subsequently, tissue fragments were washed once with 10% (v/v) heat-inactivated fetal bovine serum (hiFBS, Biowest) in HBSS to inactivate trypsin and twice with HBSS to remove FBS from the solution. Tissue fragments were then mechanically dissociated with a 5 ml plastic pipette and subsequently with 1 ml pipette tips. Viable cells were counted using the trypan blue (0.4%

(w/v), Sigma-Aldrich Co.) exclusion assay, and the cell density was adjusted to $2 \times 10^7$ viable cells/ml. Thereafter, 5 μl of the cell suspension was seeded in the cell body compartment of a μEF device, previously treated with 0.01 mg/ml PDL as described above. Cells were cultured in Neurobasal medium supplemented with 0.5 mM glutamine, 2% B27, and 1% penicillin/streptomycin (P/S, 10,000 units/ml penicillin and 10,000 μg/ml streptomycin) and kept in a humidified incubator at 37 °C supplied with 5% $CO_2$.

Primary embryonic mouse dorsal root ganglion (DRG) explants were isolated from wild-type C57BL/6 mice embryos (E16.5). Lumbar DRG explants were removed and placed in HBSS until use. Upon use, one DRG explant was seeded in the cell body compartment of a μEF device, previously treated with 0.01 mg/ml PDL plus laminin as described above. Cells were cultured in Neurobasal medium supplemented with 0.5 mM glutamine, 2% B27, 50 ng/ml of NGF 7S (Calbiochem®, Millipore), and 1% P/S, and kept in a humidified incubator at 37 °C supplied with 5% $CO_2$.

**Electrophysiology recordings**

Cortical neurons at day *in vitro* (DIV) 15 and DRG explants at DIVs 4, 6, and 8, were used in the electrophysiology experiments. The cortical and DRG recordings were respectively obtained with 252- and 120-electrode MEAs using MEA2100 recording systems (MultiChannel Systems MCS GmbH, Germany). The μEF devices prepared with these MEAs had 16 and 12 microchannels with four and five microelectrodes positioned along each microchannel, respectively. The electrodes are referred to using the electrode labels defined by MultiChannel Systems, with a number representing the row and a letter representing the column, and the microchannels are referred to using the letter representing the electrode column over which they are aligned (see Fig. 1, low left corner of the GUI, for a schematic example of this labeling system). Recordings were obtained at a sampling rate of 20 kHz, and all recorded activity was spontaneous activity.

**Computational algorithms**

μSpikeHunter was developed in MATLAB R2016b (The MathWorks Inc., USA). The graphical user interfaces were developed using MATLAB's GUI development environment (GUIDE). With the GUIDE Layout Editor, GUI layouts were developed for compatibility with both Windows and Macintosh operating systems.

*Propagation sequence detection*

Propagation sequences are detected by first applying event detection to the $i$th electrode closest to the center of the microchannel and using a set of three criteria to determine whether each detected event is part of a propagation sequence. The user defines the number of standard deviations to use for the event detection threshold and whether the positive or negative phase of the spikes is being detected. The event detector first eliminates outliers and then calculates the median and standard deviation of the noise. Outliers are eliminated by excluding any data points that are more than three scaled median absolute deviations (MADs) from the median of the signal. The scaled MAD is calculated as

$$MAD_{sc} = C \text{ median } [S_i - \text{median }(S)]$$

where $S$ is the discrete recorded signal and $C = -1/[\sqrt{2}\,\text{erfc}^{-1}(3/2)]$ is a scaling constant. With the outliers excluded, the median $M_n$ and standard deviation $\sigma_n$ of the remaining data are obtained; these are considered to be the median and standard deviation of the noise. The detection threshold $V_D$ is calculated as the user-defined number $n_\sigma$ of standard deviations above or below the median, as

$$V_D = M_n \pm n_\sigma \sigma_n$$

Events are obtained as regions of the signal that exceed the detection threshold. The time of the $p$th event on the $i$th electrode is denoted $t_{p,i}$.

For an event detected on the central $i$th electrode to be considered part of a propagation sequence, all of the following three conditions must be met.

(1) Temporally linked events are detected on all other electrodes.
(2) The times of the linked events on the first and last electrodes correspond to a propagation velocity of less than 100 m/s.
(3) The absolute value of the Kendall rank coefficient $\tau_b$ between the electrode indices and the times of the linked events is greater than 0.8.

For condition (1), a search window $[T_{s,\min}, T_{s,\max}]$ is defined for each electrode based on the distance to the first electrode as

$$T_{s,\min} = t_{p,1} - C_s D_{1,i} \qquad T_{s,\max} = t_{p,1} + C_s D_{1,i}$$

where $D_{1,i}$ is the distance from the $i$th electrode to the first electrode and $C_s$ is a search window coefficient defined to correspond to a minimum propagation velocity of 0.1 m/s. The event detection thresholds are also obtained for all of the electrodes using the two-step procedure described previously. A collection of temporally linked events is considered to exist only if an event is detected on each electrode within each corresponding search window. The times of the $p$th propagation sequence are given by the vector $\mathbf{t}_p = [t_{p,1}\ t_{p,2}\ ...\ t_{p,n_E}]$, where $n_E$ is the number of electrodes.

For condition (2), the times of the events are defined as the times at which the peak voltages are recorded. A sequence is rejected if the following condition is not met:

$$\frac{D_{1,n_E}}{|t_{p,n_E} - t_{p,1}|} < 100 \text{ m/s}$$

In condition (3), the Kendall rank coefficient $\tau_b$ is an index of the ordinal association between two sets of numbers. For this condition, $\tau_b$ is obtained for the correlation between a vector of the electrode indices, given by $\mathbf{E} = [1\ 2\ ...\ n_E]$, and the $p$th propagation sequence time vector $\mathbf{t}_p = [t_{p,1}\ t_{p,2}\ ...\ t_{p,n_E}]$, as

$$\tau_b = \frac{2\sum_{i=1}^{n_E-1}\sum_{j=i+1}^{n_E} \xi^*(E_i, E_j, t_{p,i}, t_{p,j})}{n_E(n_E - 1)}$$

where

$$\xi^*(E_i, E_j, t_{p,i}, t_{p,j}) = \begin{cases} 1, & (E_i - E_j)(t_{p,i} - t_{p,j}) > 0 \\ 0, & (E_i - E_j)(t_{p,i} - t_{p,j}) = 0 \\ -1, & (E_i - E_j)(t_{p,i} - t_{p,j}) < 0 \end{cases}$$

A sequence is rejected if the following condition is not met:

$$|\tau_b| > 0.8$$

All as-detected propagation sequences are presented to the user in a list, and the user may view the voltage traces for each propagation sequence in a plot in the main GUI.

*Single-sequence propagation velocity*
The cross-correlations of the voltage traces for each possible pair of electrodes are calculated and plotted in the main GUI. The plotted cross-correlations $\mathbf{X}_{i,j}$ are normalized such that all autocorrelations take a unit value at zero lag, as

$$\mathbf{X}_{i,j}(\tau) = \frac{\mathbf{V}_{p,j} \otimes \mathbf{V}_{p,i}(\tau)}{\mathbf{V}_{p,i} \otimes \mathbf{V}_{p,i}(\tau)} \quad \text{where } j > i.$$

Here, $\mathbf{V}_{p,i}$ and $\mathbf{V}_{p,j}$ are the voltage traces of the $p$th propagation sequence on the $i$th and $j$th electrodes, respectively; $\tau$ is the lag time between the voltage traces; and $\otimes$ represents the cross-correlation, which is calculated as a function of $\tau$. Additionally, the condition $j > i$ is defined so that the propagation velocity estimate described below yields positive and negative velocities for anterograde and retrograde propagation, respectively. The time window for this cross-correlation is defined as $t_{p,i} \pm WD_{i,j}$, where the time window constant $W$ is 7.5 s/m.

A change of variable is also performed for the lag timescale of each cross-correlation so that the cross-correlation is given as a function of the inverse of the velocity, as

$$v_\tau^{-1} = \frac{\tau}{D_{i,j}}$$

where $v_\tau$ is the velocity a wave would have to travel to have a lag time of $\tau$ between the $i$th and $j$th electrodes separated by a distance of $D_{i,j}$. On the basis of this change of variable, the cross-correlation time window corresponds to a minimum speed of 0.067 m/s.

The time between spikes being recorded on two different electrodes can be obtained as the lag $\tau_{\text{peak}}$ at which the cross-correlation of their voltage traces is maximized, as

$$\tau_{\text{peak}} = \arg\max_\tau \left( \mathbf{X}_{i,j}(\tau) \right)$$

From this lag time and the distance between the two electrodes, the single-sequence propagation velocity (SPV) is calculated as $v_S = D_{i,j}/\tau_{\text{peak}} = v_{\tau,\text{peak}}$. The SPV can be obtained for any pair of electrodes or as the average of the SPV estimates for all pairs.

However, it should be noted that the time resolution affects the accuracy of the SPV estimates in a nonlinear way and this is not taken into consideration when the average is computed. That is, when the lag is smaller, as for more closely located electrodes, small errors in the lag produce larger errors in the SPV. Thus, the SPV is less error prone when a more distant pair of electrodes are selected. Additionally, regardless of the inter-electrode distance, an underestimation of the absolute value of the lag produces a larger error than an overestimation of the same magnitude, though the difference between the errors is larger for closer electrodes as a result of the $\tau^{-1}$ dependence. Because of this, the SPV tends to overestimate the propagation velocity, especially when a closer electrode pair or the average of all pairs is selected for the estimation.

The SPV is paired with a confidence index, which indicates the similarity of the spike shapes recorded on the two electrodes. When a single pair is selected for SPV calculation, the confidence index $CI_S$ is the peak value of the normalized cross-correlation:

$$CI_S = \max_{\tau}(\mathbf{X}_{i,j}(\tau))$$

When the average of all pairs is selected, the confidence index is the peak value that is lowest among the electrode pairs.

*Kymograph and audio playback*

µSpikeHunter also contains two interactive elements for the detection of traveling signals: a kymograph and an audio playback function. The kymograph is an image representation of the voltage signals recorded on each electrode. The voltages are converted to pixel color map intensities and plotted in time–electrode space (see Fig. 1). This representation allows the user to readily visually assess whether there is propagation, determine the direction and speed of propagation, and observe the relative peak voltage magnitudes on each electrode. The user may draw a line on the kymograph to manually calculate the propagation velocity.

The audio playback function assigns a different tonal frequency (note) to each electrode selected for analysis and converts the voltage signals on each electrode to audio intensities. The time dilation for the playback is 500 times to allow the spikes recorded on the different electrodes to be distinguishable by the human ear. With this playback function, traveling waves are detectable as a sequence of ascending or descending frequencies for anterograde or retrograde propagation, respectively.

Together with the detailed quantitative methods, these two functions allow the users to qualitatively detect traveling waves in two different modalities: visual and auditory.

*Spike sorting*

Spike sorting is performed based on regions of interest (ROIs) drawn by the user for up to four source clusters (clusters 1–4) with a fifth cluster (cluster 0) comprising any spikes not sorted into clusters 1–4. ROIs are drawn on the plot for the electrode selected in the main GUI prior to the spike sorting process; this electrode is hereafter referred to as the "event electrode." Up to two ROIs can be drawn for each cluster. Spikes are sorted sequentially from cluster 1 to 4 and are removed from the sorting pool once they have been assigned to a source cluster. This means that ROIs drawn for clusters with higher cluster identification numbers (IDs) may be drawn less selectively than for clusters with lower cluster IDs.

Because the plotted events for each electrode are all part of propagation sequences, each event on the event electrode is tied to corresponding events on all other electrodes. Thus, once the spikes are sorted on the event electrode, the spikes are also sorted on all other electrodes in accordance with the propagation sequence to which they belong. The user may then visually confirm that the intra-cluster spike shapes are consistent not only on the event electrode but also on all other electrodes.

*Cluster propagation velocity*

The cluster propagation velocity (CPV) estimate is based on the timing of the cluster voltage peaks and is calculated as follows. First, for each cluster, the events on each electrode are realigned based on the cross-correlation with every other event on the same electrode and in the same cluster. The time window of this cross-correlation is from 1.0 ms before to 1.0 ms after the time at which the peak voltage value is reached in each event. This realignment is then tied to a meaningful cluster-based feature by determining the time at which the mean of all the spikes in the cluster reaches a peak value, and the realignment times $t_{p,i}^*$ of each of the spikes are defined to correspond to this peak time. An example of this is shown in (Supplementary Material) Fig. 9. Figure 9a shows a plot of 13 events in the same cluster aligned about their minima, and Fig. 9b shows the events realigned about their realignment times $t_{p,i}^*$.

From these realignment times, the CPV of the $p$th event is calculated as

$$v_{C\,p} = \frac{t_{p,j}^* - t_{p,i}^*}{D_{i,j}} \quad \text{where } j > i$$

Here, the condition $j > i$ is defined such that anterograde and retrograde propagation yield positive and negative CPVs, respectively. The user may select any pair of electrodes to calculate the CPV; however, errors in the CPV due to the time resolution tend to be larger and more nonlinear for closer electrodes.

As with the SPV, the CPV is also accompanied by a confidence index, which indicates the similarity of the spike shapes recorded on the two selected electrodes $i$ and $j$ with all other spikes on the same electrode in the same cluster. This is represented by

$$CI_{C\,p} = \min(CI_{C\,p,i}, CI_{C\,p,j})$$

$$CI_{C\,p,i} = \frac{1}{n_{ev} - 1} \sum_{q=1}^{n_{ev}} \max_{\tau} \left[ \frac{\mathbf{V}_{p,i} \otimes \mathbf{V}_{q,i}(\tau)}{\mathbf{V}_{p,i} \otimes \mathbf{V}_{p,i}(\tau)} \right] \quad \text{for } q \neq p$$

where the cross-correlation is performed over the same time window as stated previously (from 1.0 ms before to 1.0 ms after the peak voltage in each event).

It should be noted that the SPV and CPV estimates of the propagation velocity are fundamentally different in their approaches. Whereas the SPV estimates the lag for a single propagation sequence by matching the spike waveforms detected on two electrodes, the CPV essentially estimates the delay between the peak voltages on two electrodes based on the timing of the peak voltage of a "master spike" representing the cluster. Therefore, when the waveforms are not consistent across all electrodes, the SPV and CPV can yield different results, as the lag that yields the highest cross-correlation for the SPV may not correspond well to the lag between the peak voltages on the two considered electrodes.

**Generation of synthetic data for validation**

Synthetic data were generated in MATLAB R2016b to validate the propagation sequence detection and propagation velocity estimation capabilities of µSpikeHunter. Synthetic spikes were generated as one phase of a sinusoidal wave with an amplitude of $V_{\text{peak}} = 60$ µV and a duration of $t_s = 1.5$ ms. Four synthetic voltage traces $V_i$ ($i = 1,2,3,4$) were generated with a sampling rate of $f = 20$ kHz to represent a series of four electrodes with an inter-electrode spacing of 100 µm. The delays between the spikes comprising a propagation sequence were defined to correspond to a propagation velocity of 0.5 m/s. The inter-spike interval was set to 25 ms, and the duration of each recording dataset was defined based on the added noise to yield approximately 40–70 detectable propagation sequences in each dataset.

Signals were generated with different levels of additive noise, and the effect of the noise on the performance of µSpikeHunter was analyzed. The noise added to the signals was defined to have a "memory" equal to the duration of the spikes (1.5 ms). That is, first, for each voltage trace, a vector $R$ of random values drawn from a normal distribution ($\mu = 0$, $\sigma = 1$) was generated. Each element of the additive noise vector was then generated by summing the previous $s$ values of the random value vector and normalizing by $s$, as

$$V_{\text{noise},i} = \frac{V_{\text{peak}}}{SNR} R \otimes k$$

where $s$ ($= f t_s$) is the number of samples spanning a spike, $V_{\text{noise},i}$ is the noise signal to be added to the $i$th voltage trace $V_i$, $SNR$ is the signal-to-noise ratio (SNR) defined as the ratio of $V_{\text{peak}}$ to the maximum possible value of $V_{\text{noise},i}$, and $k$ is a one-dimensional convolution kernel of length $s$ where every element is equal to $1/s$.

The considered SNRs ranged from 0.2 to 0.7 in intervals of 0.1, and for each SNR, three recording datasets were created. At each SNR, the propagation sequence detection performance and the accuracy of the different propagation velocity estimates were evaluated. The analysis of the synthetic data in µSpikeHunter presented in the validation results below was conducted with a user-input event detection threshold of 2.2 standard deviations. This threshold was selected to yield sufficiently few detected events when a dataset containing only noise on the four simulated electrodes was loaded into µSpikeHunter. Three noise datasets of 100 s were analyzed, and each yielded fewer than 5 events with an average of 1.3 events per dataset.

**Data availability statement**
The µSpikeHunter software, as well as a sample test dataset, is available for download from [SourceForge/Github links].

**Acknowledgments/Funding**
This work was partly financed by FEDER - Fundo Europeu de Desenvolvimento Regional funds through the COMPETE 2020 - Operational Programme for Competitiveness and Internationalisation (POCI), Portugal 2020, and by Portuguese funds through FCT - Fundação para a Ciência e a Tecnologia/ Ministério da Ciência, Tecnologia e Inovação in the framework of the project "Institute for Research and Innovation in Health Sciences" (POCI-01-0145-FEDER-007274) and in the framework of the financed project PTDC/BIMMED/4041/2014. Paulo Aguiar was supported by Programa Ciência – Programa Operacional Potencial Humano (POPH) – Promotion of Scientific Employment, ESF and MCTES and program Investigador FCT, POPH and Fundo Social Europeu. The microfluidic chambers used in this study were fabricated at INESC - Microsystems and Nanotechnologies, Portugal, under the supervision of João Pedro Conde and Virginia Chu.

**Author Contributions Statement**
P.A. devised the project and main conceptual ideas. K.H. and P.A. developed the algorithms, analyzed the data and wrote the main manuscript text. K.H. worked out almost all of MATLAB code implementation. J.M., C.L. and E.N. carried out the neuronal culture experiments. M.L. was responsible for the co-culture application ideas. All authors discussed the results and contributed to the final manuscript.

**Competing interests**
The authors declare no competing interests.

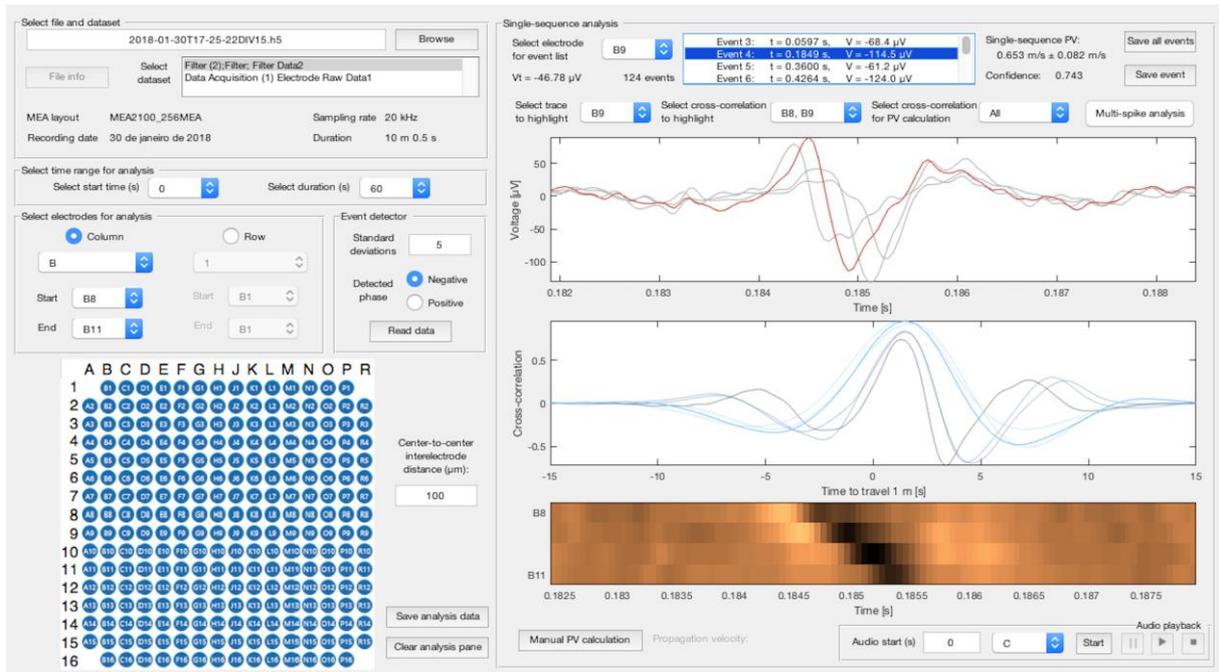

Figure 1. Main graphical user interface (GUI) of µSpikeHunter.

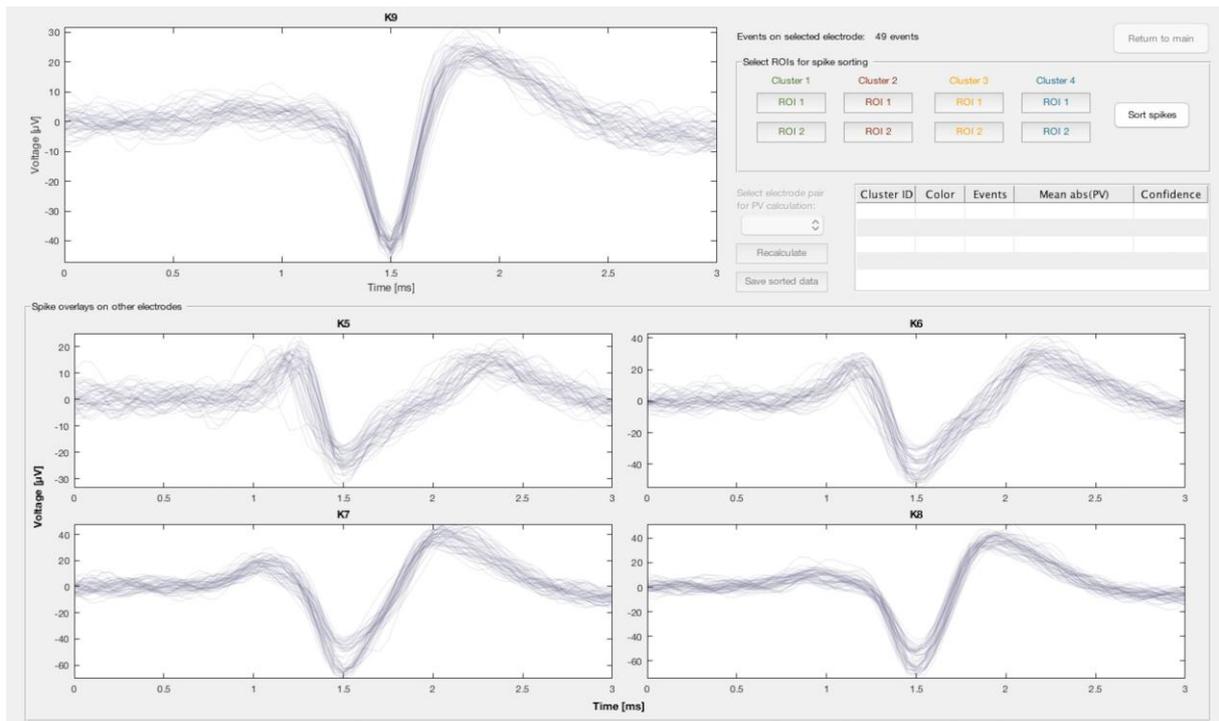

Figure 2. Spike sorting graphical user interface (GUI) of µSpikeHunter.

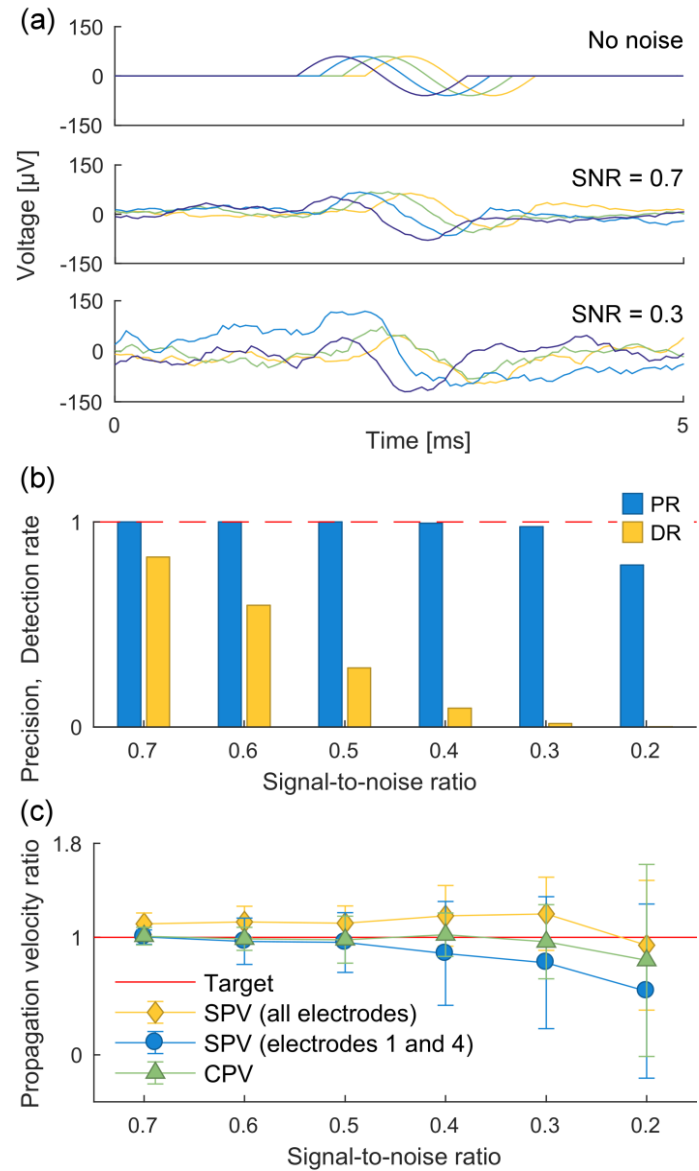

Figure 3. Validation results. (a) Examples of synthetic propagation sequences detected by the sequence detector in the case of no noise and SNRs of 0.7 and 0.3. (b) Precision and detection rate of propagation sequence detector at different SNRs. The dashed line indicates the target value of 1.0. (c) Ratios of different propagation velocity estimates to the true propagation velocity at different SNRs. Estimates were obtained for three recording datasets with approximately 40–70 sequences each at each SNR and averaged over the three datasets. Error bars represent the standard deviations.

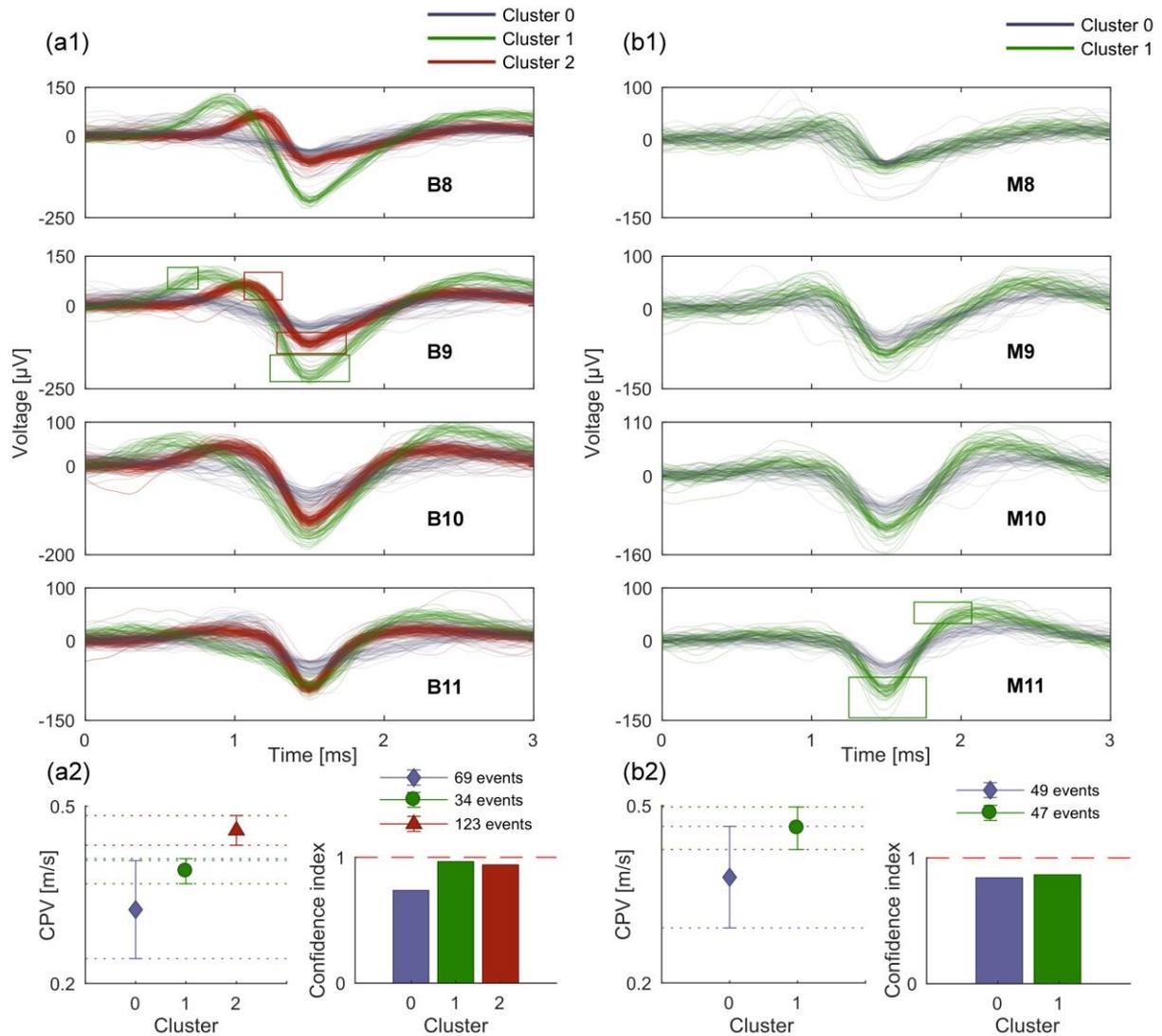

Figure 4. Examples of μSpikeHunter spike sorting results for spikes recorded from cortical neurons at DIV 15 using the MEA2100-256-System with a sampling rate of 20 kHz and an analysis period of 100 s. (a1) Microchannel B with three source clusters. The boxes on the B9 plot are the ROIs used for sorting. (a2) Intra-cluster averages of the CPV and confidence index for channel B events and the numbers of events in each cluster. (b1) Microchannel M with two source clusters. The boxes on the M11 plot are the ROIs used for sorting. (b2) Intra-cluster averages of the CPV and confidence index for channel M events and the numbers of events in each cluster.

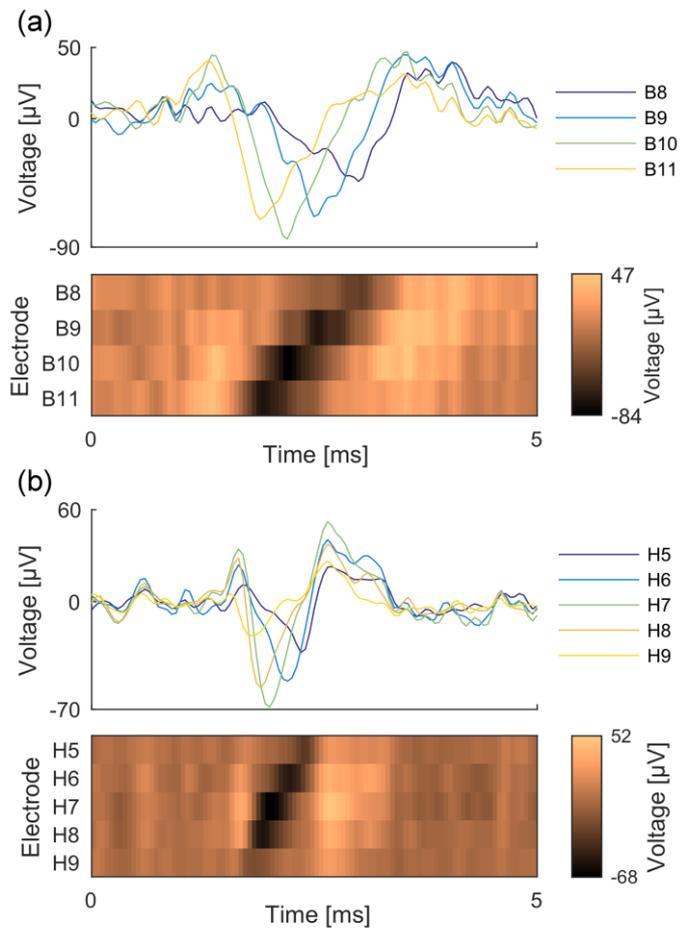

Figure 5. Examples of voltage traces and kymographs of reverse propagating APs recorded from (a) cortical and (b) DRG cultures.

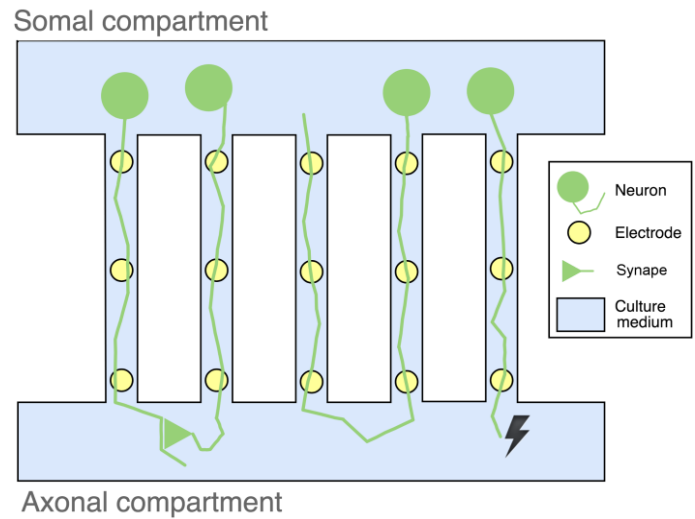

Figure 6. Schematic of possible causes of reverse propagation. Left two channels: signal transmission via an excitatory axo-axonal synapse. Middle two channels: an axon growing back through a microchannel to the somal compartment. Rightmost channel: signal generation from the axon terminal (spontaneous or induced) followed by antidromic conduction.

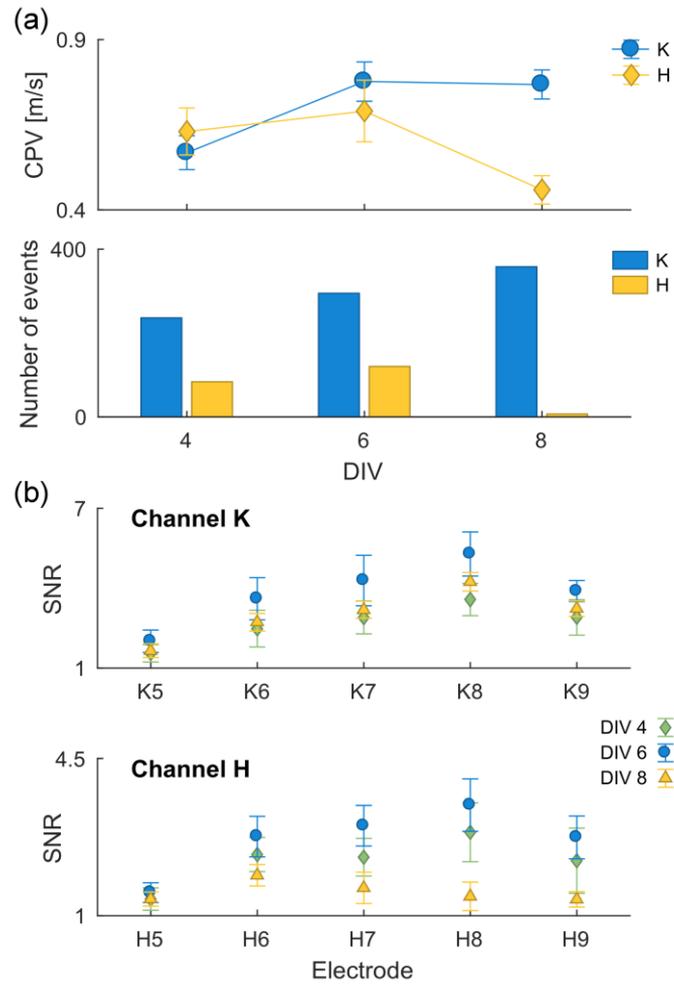

Figure 7. Monitoring DRG culture activity over DIVs 4, 6, and 8, using the MEA2100-120-System with a sampling rate of 20 kHz and a recording duration of 10 min. Sample results for microchannels channels K and H. (a) Propagation velocity and number of spikes. (b) SNR for each electrode in the two presented microchannels.

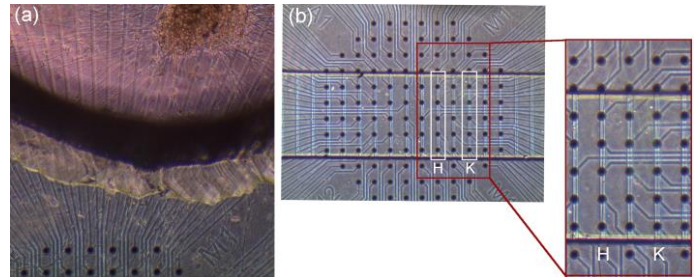

Figure 8. Microscope images of a dorsal root ganglion (DRG) explant culture in a µEF device. (a) The DRG explant in the somal compartment is visible at the top of the image. (b) Alignment of the microchannels over the microelectrodes. The inset shows a magnified view of specific microgrooves/channels, H and K.

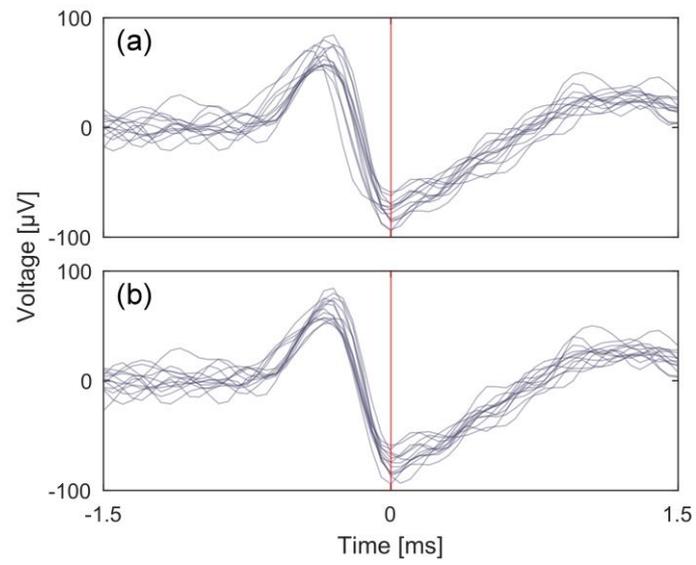

Figure 9. Demonstration of spike realignment based on the intra-cluster cross-correlation. (a) Spikes aligned about their minima. (b) Spikes realigned based on the cross-correlation.